\documentclass[12pt]{article}
\usepackage[latin1]{inputenc}
\usepackage{amsmath}
\usepackage{amsfonts}
\usepackage{amssymb}
\usepackage{pstricks}
\usepackage{amsthm}
\usepackage{mathrsfs}
\usepackage{amssymb}
\usepackage{cancel}
\usepackage{slashed}
\usepackage{graphicx}
\usepackage[font=small,labelfont=bf]{caption}
\usepackage{float}
\usepackage{csquotes}
\usepackage[bottom]{footmisc} 
\usepackage{tikz}
\usepackage{setspace}
\usepackage{bbding}
\usepackage{cite}
\usetikzlibrary{matrix,arrows,decorations.pathmorphing}
\usepackage{framed}
\usepackage{color}
\usepackage{wrapfig}
\definecolor{shadecolor}{RGB}{224,238,238}
\newcommand{\nn}{\nonumber}

\makeatletter

\@addtoreset{equation}{section}
\makeatother

\def\lsim{\;\raise0.3ex\hbox{$<$\kern-0.75em\raise-1.1ex\hbox{$\sim$}}\;}
\def\gsim{\;\raise0.3ex\hbox{$>$\kern-0.75em\raise-1.1ex\hbox{$\sim$}}\;}
\def\beq{\begin{equation}}   \def\eeq{\end{equation}}
\def\ba{\begin{array}}       \def\ea{\end{array}}
\def\bea{\begin{eqnarray}}   \def\eea{\end{eqnarray}}
\def\nn{\nonumber}
\def\nl{\newline}

\def\ve{\varepsilon}

\theoremstyle{definition} 
\date{\today}

\begin{document}

\begin{titlepage}
\begin{flushright}
LPT Orsay 16-54 
\end{flushright}


\begin{center}

\begin{doublespace}

\vspace{1cm}
{\Large\bf Possible Explanation of the Electron Positron Anomaly at 17~MeV
in $^8Be$ Transitions Through a Light Pseudoscalar} \\
\vspace{2cm}

{\bf{Ulrich Ellwanger$^{a,b}$ and Stefano Moretti$^b$}}\\
\vspace{1cm}
{\it  $^a$ Laboratoire de Physique Th\'eorique, UMR 8627, CNRS, Universit\'e de Paris-Sud,\\
Univ. Paris-Saclay, 91405 Orsay, France\\
\it $^b$ School of Physics and Astronomy, University of Southampton,\\
\it Highfield, Southampton SO17 1BJ, UK}

\end{doublespace}

\end{center}
\vspace*{2cm}

\begin{abstract}
We estimate the values of Yukawa couplings of a light pseudoscalar $A$ with a
mass of about 17~MeV, which would explain the $^8Be$ anomaly observed in
the Atomki pair spectrometer experiment. The resulting couplings of $A$ to up and
down type quarks are about 0.3 times the coupling of the standard Higgs boson.
Then constraints from $K$ and $B$ decays require that loop contributions to flavour
changing vertices cancel at least at the 10\% level. Constraints from beam
dump experiments require the coupling of $A$ to electrons to be larger than
about 4 times the coupling of the standard Higgs boson, leading to a short enough
$A$ life time consistent with an explanation of the anomaly.
\end{abstract}

\end{titlepage}

\newpage

\section{Introduction}

The Atomki pair spectrometer experiment \cite{Krasznahorkay:2015iga} has searched
for electron-positron internal pair creation in the decay of excited $^8Be$ nuclei. The $^8Be$
excitations were produced with help of a beam of protons directed on a $^7Li$ target
and the different $^8Be$ excitations could be separated by tuning the energy of the
incoming protons.

An anomaly has been observed in the decay of $^8{Be^*}$ with
spin-parity $J^P=1^+$ into the ground state $^8Be$ with spin-parity $0^+$ (both with isospin $T=0$),
where $^8{Be^*}$ has an excitation energy of 18.15~MeV.
Both distributions of the opening
angle $\theta$ of the electron-positron pair and the invariant mass of the
electron-positron pair showed an excess consistent with an intermediate boson $X$
being produced in the decay of $^8{Be^*}$, with $X$ decaying into an electron-positron pair.
The best fit to the mass $M_X$ of $X$ is \cite{Krasznahorkay:2015iga}
\beq
\label{fitMX}
M_X = 16.7 \pm 0.35\ \text{(stat)}\ \pm 0.5\ \text{(sys)\ MeV}
\eeq
whereas the best fit to the branching fraction $^8{Be^*} \to {^8Be}+(X\to e^+ e^-)  $ relative
to the branching fraction $^8{Be^*} \to  {^8Be}+ \gamma$ is given by
\beq
\label{fitBR}
\frac {Br(^8{Be^*} \to X + {^8Be})\times Br(X\to e^+ e^-)}
{Br(^8{Be^*} \to \gamma + {^8Be})} = 5.8 \times 10^{-6}\; .
\eeq
These values correspond to a statistical significance of the excess of $6.8\,\sigma$
 \cite{Krasznahorkay:2015iga}.

In the case of the excitation $^8{Be^*}'$ with spin-parity $1^+$ (but isospin $T=1$)
and an excitation energy of 17.64~MeV, no excess was observed. The simplest
explanation is that this decay is kinematically suppressed; this kinematical
suppression is the stronger the heavier the intermediate boson $X$ would be. This
motivates a value of $M_X$ somewhat above the best fit value in \eqref{fitMX}
(which may lead to a somewhat smaller statistical significance and smaller
best fit to the relative branching fraction).

In \cite{Feng:2016jff,Feng:2016ysn} an explanation for the observed excess was
given in the form of models featuring a new vector boson $Z'_\mu$ with a mass
$M_{Z'}$ of about 17~MeV, with vector-like couplings to quarks and leptons.
Constraints on such a new vector boson, notably from searches for $\pi^0\to Z'+\gamma$
by the NA48/2 experiment \cite{Batley:2015lha}, require that the couplings
of $Z'_\mu$ to up and down quarks are ``protophobic'', i.e., that the charges
$e \ve_u$ and $e \ve_d$ of up and down quarks -- written as multiples of the
positron charge $e$ -- satisfy $2\ve_u+\ve_d \lsim 10^{-3}$
\cite{Feng:2016jff,Feng:2016ysn}. Subsequently, further studies of such models
have been performed in \cite{Gu:2016ege,Chen:2016dhm,Liang:2016ffe,Jia:2016uxs}.

Given the quantum numbers of the $^8{Be^*}$ and ${^8Be}$ states, the boson $X$ can
also be a pseudoscalar $A$ with a mass $M_A$ of about 17~MeV. In
\cite{Feng:2016jff,Feng:2016ysn} this possibility is dismissed quite rapidly.
The argument is that, for such an axion-like pseudoscalars $A$, fermion loops generate
couplings of the form $g_{A\gamma\gamma} A F^{\mu\nu}(\gamma)\tilde{F}_{\mu\nu}(\gamma)$
which are strongly constrained by axion searches. However, light pseudoscalars
in this mass range
with tree level Yukawa couplings to electrons decay dominantly into electron-positron
pairs, unless Yukawa couplings to other charged fermions $f$ with mass $m_f$ are much larger
than $m_f/m_e$ compensating $g_{A\gamma\gamma}\approx 1/(8\pi m_f)$.

It is the purpose of the present paper to study the required couplings of a pseudoscalar
$A$ with a mass of about 17~MeV in order to explain the ${^8Be}$ anomaly observed
in \cite{Krasznahorkay:2015iga}, and to verify under which conditions these
couplings satisfy existing constraints. We have in mind a pseudoscalar $A$ originating
from extended Higgs sectors of the Standard Model (SM) including, e.g., two Higgs
doublets of type II and a singlet as in the Next-to-Minimal Supersymmetric SM
(NMSSM) \cite{Ellwanger:2009dp},
where $A$ could be very light in Peccei-Quinn or $R$-symmetry limits \cite{Ellwanger:2009dp}.
We find however that (singlet extended) two Higgs doublet models of type II have
difficulties to explain
the anomaly, but more general models are possible under the condition that
the various loop contributions to the flavour changing vertex $A-s-d$ cancel
at least at the 10\%  level.

A major task is to express the coupling of such a pseudoscalar to $^8{Be^*}$ and ${^8Be}$
states in terms of the couplings of $A$ to up and down quarks. Required is
actually the ratio of branching fractions
\beq
\label{BRAtoGamma}
\frac {Br(^8{Be^*} \to A + {^8Be})\times Br(A\to e^+ e^-)}
{Br(^8{Be^*} \to \gamma + {^8Be})}
\eeq
which is given in \eqref{fitBR}. In the case of the $Z'$ considered in
\cite{Feng:2016jff,Feng:2016ysn}, use is made of the fact that both $Z'$ and
photons couple via conserved currents to quarks, an argument which is not useful
here. Furthermore, \cite{Feng:2016jff,Feng:2016ysn} argue that both $Z'_\mu$ and
photons couple via conserved currents to nucleons, and that -- at least in the
isospin conserving limit considered in \cite{Feng:2016jff} -- matrix elements of
conserved currents cancel in the calculation of the ratio of decay widths
up to the modifications of the couplings.
(The possible impact of isospin violating effects is analysed in \cite{Feng:2016ysn}.)

The calculation of the coupling of a pseudoscalar $A$ to $^8{Be^*}$ and ${^8Be}$ states 
has to proceed in two steps. Firstly, the couplings of $A$ to nucleons have to
be obtained: These are proportional to the nucleon quark spin components $\Delta q$,
and have been studied in the context of direct detection of dark matter via
the exchange of pseudoscalars, e.g., in \cite{Cheng:2012qr,Dolan:2014ska}.
Secondly, the $^8{Be^*}$ and ${^8Be}$ nuclei have to be described in terms of
nucleons with definite spin, angular momentum and total momentum. To this end
we employ wave functions from the simple unperturbed nuclear shell model.
We are aware of the fact that this approach is somewhat simplistic: It neglects 
proton-neutron pairing effects, $\alpha-\alpha$ substructures of the ${^8Be}$ states
and, in particular, possible mixing with the nearby $^8{Be^*}'$ state induced by
isospin breaking. Effects of the latter have been discussed in \cite{Feng:2016ysn},
and could be sizeable.
For consistency, we have to employ the same approach for the decay widths
$\Gamma(^8{Be^*} \to \gamma + {^8Be})$ and $\Gamma(^8{Be^*} \to A + {^8Be})$.
One may hope that the inaccuracies of the nuclear shell model wave functions cancel to some
extent in the calculations of the ratio of decay widths, but we will return to
this issue later on. In any case some theoretical error has certainly
to be taken into account, and a further refinement of the present calculation
of this ratio would be desirable.

The plan of the paper is as follows.
In section~2 we consider the couplings of a pseudoscalar to nucleons while in section~3 we compute and compare the relevant matrix elements for $\gamma$
and pseudoscalar emission in the nuclear shell model. In this section we also find
the conditions on the pseudoscalar Yukawa couplings to quarks and leptons
which are necessary in order to explain the anomaly. Section~4 is devoted to
other experimental constraints on these couplings. Finally, a summary and some conclusions are
presented in section~5.

\section{Couplings of a pseudoscalar to nucleons}

Subsequently we define reduced couplings $\xi_q$ of a pseudoscalar $A$ to quarks
in terms of
\beq
{\cal L}_{Aqq} = \xi_q \frac{m_q}{v} A \bar{q} i \gamma_5 q
\eeq
with $v\sim 246$~GeV. As in \cite{Cheng:2012qr} we define a pseudoscalar-nucleon
coupling $h_N$ (with $N=p,n$ for protons and neutrons, respectively) by
\beq
h_N=\frac{1}{v} \sum_q\left<N|\xi_q m_q \bar{q} i \gamma_5 q|N\right>\; .
\eeq
From \cite{Cheng:2012qr} (see also \cite{Dolan:2014ska}) one finds
\beq
h_N=\frac{m_N}{v}\sum_{q=u,d,s}\Delta_q^{(N)}\left(\xi_q -
\sum_{q'=u,...,t}\xi_{q'}\frac{\bar{m}}{m_{q}}\right)\; ,
\eeq
where $\Delta_q^{(N)}$ are the quark spin components of the nucleon $N$, and
$\bar{m}=\frac{1}{m_u^{-1}+m_d^{-1}+m_s^{-1}}\sim \frac{m_u m_d}{m_u+m_d}$.
In addition, we assume \cite{Cheng:2012qr}
$m_d\sim 2m_u\sim 2\times 2.5$~MeV and
\beq
\xi_u=\xi_c=\xi_t,\qquad \xi_d=\xi_s=\xi_b\; .
\eeq
Neglecting $\frac{m_{u,d}}{m_{s,c,b,t}}$ one obtains
\beq
h_N = \frac{m_N}{v}\left(\Delta_u^{(N)}(-\xi_u-2\xi_d)+\Delta_d^{(N)}(-\xi_u)
+\Delta_s^{(N)}\xi_d\right)\; .
\eeq
For $\Delta_q$ we use the values given in Table~II in \cite{Cheng:2012qr} using
$g^8_A = 0.46$ and $g^0_A = 0.37$:
\beq
\Delta_u^{(p)}=0.84,\ \Delta_d^{(p)}=-0.44,\ \Delta_s^{(p)}=-0.03,\
\Delta_u^{(n)}=-0.44,\ \Delta_d^{(n)}=0.84,\ \Delta_s^{(n)}=-0.03\; .
\eeq
This gives
\beq
h_p=\frac{m_p}{v}\left(-0.40\xi_u-1.71\xi_d\right),\quad
h_n=\frac{m_n}{v}\left(-0.40\xi_u+0.85\xi_d\right)\; .
\eeq
For the average $\bar{h}_N^2 \equiv \frac{(h_p+h_n)^2}{4}$, required for the
$^8{Be^*}$ decays, one obtains (with $m_n \sim m_p$)
\beq
\label{fxi}
\bar{h}_N^2 = \frac{m_p^2}{v^2}f(\xi_u,\xi_d), \qquad
f(\xi_u,\xi_d) =
\left(0.16\xi_u^2+0.35\xi_u\xi_d+0.19\xi_d^2\right) \; .
\eeq
For $\xi_u=\xi_d\equiv \xi$ one has $f(\xi,\xi)\sim 0.7\,\xi^2$.

\section{Nuclear shell model and emission matrix elements}

The $^8Be$ ground state with $J^P=0^+$ and the $^8Be^*$ excited state with
$J^P=1^+$ can be described in terms of the lowest two shells of the nuclear
shell model: The lowest 1s ($L=0$) shell is fully occupied by two nucleons
with spin $S_z=\pm 1/2$ (two out of the four protons and two out of the four
neutrons); in the next 1p ($L=1$) shell there is, a priori, space for six nucleons
with angular momentum $L_z=-1,0,+1$ and $S_z=\pm 1/2$, respectively.
However, the spin-orbit interaction proportional to $-\left<\vec{L}\cdot\vec{S}\right>$
splits the 1p level into two levels with total angular momentum $J=3/2$ 
(four possible states $1p_{3/2}$) and $J=1/2$ (two possible states $1p_{1/2}$)
where the $J=3/2$ level is lower. In the $^8Be$ ground state two out of the four $1p_{3/2}$ states
are occupied by protons/neutrons respectively, and the angular momenta can
be combined pairwise to form a nucleus with $J^P=0^+$.

If one of the two states in the lower $1p_{3/2}$ level is lifted into the previously
empty $1p_{1/2}$ level it would form with its remaining partner in the $1p_{3/2}$ level
a $J^P=1^+$ state which gives, together with the remaining $J^P=0^+$ nucleons, a
$J^P=1^+$ state consistent with the quantum numbers of $^8Be^*$. Its excitation
energy of 18.15~MeV is consistent with -- following \cite{Fayache:1996hu}
perhaps slightly larger than -- the expectations from nuclear spin-orbit splitting.
During the transition from $^8Be^*$ to $^8Be$ a photon or -- as considered here --
a pseudoscalar can be emitted emitted from a single nucleon falling from a $1p_{1/2}$
state into the lower $1p_{3/2}$ state. The photon emission is of the M1 type.

The next task is to construct the interaction Hamiltonian for both M1 photon
and pseudo\-scalar emissions from single $1p_{1/2}$ nucleon states; finally we need the
ratios of both decay rates which should be compared -- together with the
$A\to e^+ e^-$ branching fraction -- to $5.8\times 10^{-6}$ \eqref{fitBR}
as estimated for the signal in \cite{Krasznahorkay:2015iga}.

In order to treat the photon and pseudoscalar emissions at the same level
we construct first the non-relativistic interaction Hamiltonian from the relativistic
Dirac equation for single nucleons $N=p,n$.
After adding a coupling $h_N$ to a pseudoscalar $A$ and an anomalous magnetic moment
$\sim (g-2)_N$ to the Lagrangian, the Dirac
equation including the covariant U(1)$_{em}$ derivative with a photon $A^\mu=(\phi,A^i)$
can be written as (isolating the time derivative)
\beq
\label{Dirac}
i\hbar \gamma^0 \partial_t\psi_N =(\gamma^i(p^i-q_NA^i)+\gamma^0 q_N\phi
+\frac{(q\cdot (g-2))_N}{8m_N}\sigma_{\mu\nu}F^{\mu\nu}
+m_N+ih_NA\gamma^5)\psi_N
+ ...
\eeq
where the dots describe the potential (including spin-orbit terms etc.) for single
nucleons generated by the seven remaining nucleons of $^8Be$.

Decomposing $\psi_N=\left(\ba{c}\tilde{\varphi}_N\\ \tilde{\chi}_N\ea \right)$,
$\tilde{\varphi}_N=e^{im_Nt}\varphi_N$, $\tilde{\chi}_N$ can be eliminated in the non-relatistic
limit in an expansion in $1/m_N$. To lowest order in the couplings $e,\ h_N$
the remaining Schr\"odinger equation for $\varphi_N$ contains an interaction
Hamiltonian of the form
\beq
\label{Hint}
H_{int}=-q_N\vec{r}\cdot \vec{E}-\frac{1}{2m_N}\left(q_N\vec{B}\cdot \vec{L}+(q\cdot g)_N
\vec{B}\cdot \vec{S}+2h_N(i\vec{\nabla} A)\cdot \vec{S}\right)
\eeq
where the first term is irrelevant for M1 transitions, and $\vec{S}=\frac{1}{2}
\vec{\sigma}$.
In \eqref{Hint} the $g$-factor $(q\cdot g)_N$ includes the anomalous magnetic
moment $\sim (g-2)$: For protons one has to use
$q_p=e$, $g_p=5.6$, for neutrons $q_n=0$ in the first terms, but 
$(q\cdot g)_n= -3.8 e$.
The coupling of the pseudoscalar $A$ is as expected: $\vec{\nabla} A$
indicates that $A$ can be emitted only as a $p$-wave, and couples to the spin.

Next one has to evaluate the matrix elements of $H_{int}$ between the states
$\left<J'=3/2, m_{j'}\right|$ and $\left|J=1/2, m_j\right>$; the decay
rates are proportional to
\beq
\label{matr}
\sum_{m_{j'}} \left|\left<3/2, m_{j'}\right| H_{int} \left|1/2, m_j\right> \right|^2
\eeq
where one has to average over $m_j=\pm 1/2$. From the different terms in the decay
rates one can estimate the ratio between photon and pseudoscalar emission.

Let us emit the photon with momentum $\vec{p}_{\gamma}$ and the pseudoscalar with
momentum $\vec{p}_A$ in the $z$~direction, leading to
$\left|B_x\right|^2 = \left|B_y\right|^2 = \vec{p}_{\gamma}^2 \left|A_\mu\right|^2$.
Then one finds for \eqref{matr} (still for a given nucleon $N$)
\bea
\frac{1}{4m_N^2}\sum_{m_{j'}} \Big[2 B_x^2
\left|q_N \left<3/2, m_{j'}\right|  L_x \left|1/2, m_j\right>
+(q\cdot g)_N\left<3/2, m_{j'}\right|S_x \left|1/2, m_j\right> \right|^2
 \nn \\
 +
4h_N^2 \vec{p}_A^2 A^2\left|\left<3/2, m_{j'}\right|S_z  \left|1/2, m_j\right> \right|^2
\Big]\; ,
\label{mat2}
\eea
and, finally, after evaluating the matrix elements of $L_x,\ S_x$ and $S_z$,
\beq
\label{mat3}
\frac{1}{2}
\sum_{m_{j'},m_j} \left|\left<3/2, m_{j'}\right| H_{int} \left|1/2, m_j\right> \right|^2
= \frac{1}{9m_N^2}\cdot \left[ \vec{p}_{\gamma}^2 \left|A_\mu\right|^2
\left(q_N - (q\cdot g)_N\right)^2 + 2 h_N^2 \vec{p}_A^2 A^2\right]
\eeq
which, for isospin singlet nuclei, has to be averaged over the nucleon states $N=p$ and
$N=n$ (including interference terms). The two terms
$\sim \left|A_\mu\right|^2$ and $\sim A^2$ on the right hand side of \eqref{mat3}
correspond to the emission of the photon $\gamma$ and pseudoscalar $A$, respectively.
Using the expressions given below eq.~\eqref{Hint}, the average of the
coefficient $\left(q_N-(q\cdot g)_N\right)^2$ becomes\beq
\label{photaverage}
\frac{1}{4}\left(q_p-(q\cdot g)_p + q_n-(q\cdot g)_n\right)^2 \simeq 0.16\, e^2\; .
\eeq
The average for pseudoscalar couplings $2\bar{h}_N^2$ is from \eqref{fxi}
\beq
2\bar{h}_N^2 = \frac{2m_p^2}{v^2} f(\xi_u,\xi_d)
 = 2.92\times 10^{-5} f(\xi_u,\xi_d)\; .
\eeq

The decay rates also depend on powers of  the photon/pseudoscalar momenta
which originate from the phase space and normalization of the plane waves $A_\mu$
and $A$; the final dependence on the momenta is $\sim |\vec{p}|^3$ in both cases.
For the ratio of the decay rates one obtains then
\beq
\frac{Br(^8Be^* \to {^8Be} + A)}{Br(^8Be^* \to {^8Be} + \gamma)}= 
\frac{ 2.92\times 10^{-5}f(\xi_u,\xi_d)}{0.16\, e^2}  \frac{|\vec{p}_A|^3}{|\vec{p}_\gamma|^3}=
2\times 10^{-3} f(\xi_u,\xi_d)  \frac{|\vec{p}_A|^3}{|\vec{p}_\gamma|^3}\; ,
\eeq
where $e^2\simeq 0.091$ was used. Assuming a $Br(A\to e^+e^-)\sim 1$ (see below),
this expression should give
\beq
\frac{Br(^8Be^* \to {^8Be} + A)}{Br(^8Be^* \to {^8Be} + \gamma)} \approx 5.8\times 10^{-6}\; .
\eeq
The ratio of momenta depends on $M_A$. Taking $M_A=17$~MeV leads to
\beq
\frac{|\vec{p}_A|^3}{|\vec{p}_\gamma|^3} \sim 0.045\; .
\eeq
From the three previous equations one obtains
\beq
\label{feq9}
f(\xi_u,\xi_d) \stackrel{!}{\approx} 0.062\; .
\eeq
Approximating $f(\xi_u,\xi_d)$ by $f(\xi_u,\xi_d)\sim 0.175\left(\xi_u+\xi_d\right)^2$
gives
\beq
\label{xiud}
\xi_u + \xi_d \stackrel{!}{\approx} 0.6
\eeq
or, for $\xi_u=\xi_d\equiv \xi$, $\xi \stackrel{!}{\approx} 0.3$.

One should keep in mind, however, that this result depends on the use of
the nuclear shell model wave functions with definite isospin $T=0$. In particular,
the coefficient 0.16 on the right hand side of \eqref{photaverage} originates from substantial
cancellations in the case of isoscalar $M1$ transition strengths, a phenomenon
underlined before in \cite{Feng:2016ysn}. If this coefficient turns out to
be larger due to a $T=1$ component in the $^8Be^*$ wave function,
the resulting value for $f(\xi_u,\xi_d)$ in \eqref{feq9} increases
as well. Of course, the expression for $f(\xi_u,\xi_d)$ given in \eqref{fxi}
would have to be corrected as well in this case, but here no strong cancellations
occur in general. Hence the theoretical uncertainty to associate to the
result \eqref{feq9} or \eqref{xiud} points towards rather larger values for $\xi_u$ and/or $\xi_d$
required to fit the anomaly observed in the Atomki pair spectrometer experiment.

We close this section with a consideration of the $A$ width and decay length.
If $A$ has Yukawa couplings to quarks and leptons which are proportional to
the Yukawa couplings of the SM Higgs boson rescaled by generation
independent factors $\xi_d \approx \xi_u \approx \xi_e$ (or $\xi_u \ll \xi_d$),
and the Yukawa couplings to BSM fermions are not much larger than the electric
charge $e$, $A$ has a branching fraction of
about 99\% into $e^+ e^-$ and only about 1\%
into $\gamma\gamma$. Its total width is then dominated by $A\to e^+ e^-$
and given by
\beq
\Gamma(A) = \xi_e^2 \frac{m_e^2}{8\pi v^2}M_A = \xi_e^2\cdot 2.9\times 10^{-15}\ \text{GeV}
\eeq
for $M_A=17$~MeV. Its decay length is
\beq
l_A=\frac{p_A}{M_A \Gamma(A)}\; .
\eeq
For the decay $^8{Be^*}\to{^8Be}+A$ with $M(^8{Be^*})-M(^8Be)=18.15$~MeV
we obtain
\beq
l_A \sim \frac{1}{\xi_e^2}\cdot  2.5\ \text{cm}\; .
\eeq
(For $M_A=17.9$~MeV, $2\,\sigma$ above the central value in \eqref{fitMX},
we obtain $l_A \sim \frac{1}{\xi_e^2}\cdot 1.1\ \text{cm}$.)
In order to explain the observed anomaly in the Atomki pair spectrometer
experiment \cite{Krasznahorkay:2015iga}, $l_A$ should then not be much larger
than 1~cm leading to
\beq
\label{xiegt1}
\xi_e \stackrel{!}{\gsim} 1\; ,
\eeq
depending somewhat on the precise value of $M_A$.

\section{Experimental constraints}

Light pseudoscalars are subject to constraints from searches for axions or
axion-like particles. For recent summaries of constraints relevant
for light pseudoscalars decaying dominantly into $e^+ e^-$ see
\cite{Dolan:2014ska,Andreas:2010ms,Essig:2010gu,Hewett:2012ns,Dobrich:2015jyk}. 
However, since we allow for different Yukawa type couplings rescaled by $\xi_u$,
$\xi_d$ and $\xi_e$ with respect to SM Higgs couplings, at least some experimental
constraints studied therein have to be reconsidered. We note that constraints
from $\pi^0\to \gamma + X$ from the NA48/2 experiment, which play a major r\^ole
for the $Z'$ scenario \cite{Feng:2016jff,Feng:2016ysn}, do not apply here
since the decay $\pi^0\to \gamma + A$ would violate parity. Furthermore, a
light pseudoscalar cannot improve the discrepancy between the measured and the
SM value of the anomalous magnetic moment of the muon since its
contribution has the wrong sign (but is smaller in absolute value than the present discrepancy).

A first class of constraints on such pseudoscalars originates from
flavour violating meson decays, analysed recently in \cite{Dolan:2014ska}.
For $M_A\sim 17$~MeV and the range of couplings relevant here these are
the decays $K^+ \to \pi^+ + X$ (constrained by the $K_{\mu 2}$ experiment
\cite{Yamazaki:1984vg}),
$K^+ \to \pi^+ + invisible$ as measured by the experiments E787 \cite{Adler:2004hp} and
BNL-E949 \cite{Artamonov:2009sz},
$B_s\to \mu^+\mu^-$ (measured by the LHCb collaboration \cite{Aaij:2013aka}
and the CMS collaboration \cite{Chatrchyan:2013bka}, see \cite{CMS:2014xfa}
for a LHCb/CMS combination), and
$B^0\to K^0_S + invisible$ measured by CLEO \cite{Ammar:2001gi}.

Concerning $K^+ \to \pi^+ + X$, \cite{Yamazaki:1984vg} searched for an
anomalous line corresponding to $\pi^+$ in
the $K_{\mu 2}$ experiment, which would appear for
$K^+ \to \pi^+ + A$ decays independently of subsequent $A$ decays. This process depends on a
loop-induced $A-s-d$ vertex (with $W$ bosons and up-type quarks in the
loop, to be supplemented at least by $H^\pm$ bosons in consistent multi-Higgs
extensions of the SM) which depends, in turn, on the couplings of $A$
to down and up type quarks (and to $W^\pm H^\mp$). 

Constraints from Fig.~2 in \cite{Yamazaki:1984vg} have been applied to a light
pseudoscalar in the NMSSM in \cite{Andreas:2010ms}. Here squark/chargino loops
are considered, which are dominant for large $\tan\beta$ ($\xi_d \gg \xi_u$)
\cite{Hiller:2004ii}. The resulting bound
on $C_{Aff}$ in \cite{Andreas:2010ms} can be translated into $\xi_d = C_{Aff}$,
which  for $M_A \sim 17$~MeV is
\beq
\label{boundxid}
\xi_d \lsim 2\times 10^{-2}\; .
\eeq
An even stronger bound has been derived in \cite{Dolan:2014ska} in terms of $g_Y$,
a common factor rescaling the Higgs-like Yukawa couplings of $A$. Note
that $\xi_u = \xi_d \equiv \xi$ corresponds to $g_Y = \xi/ \sqrt{2}$ in
\cite{Dolan:2014ska}. These authors find that $g_Y \gsim 5\times 10^{-3}$
or $\xi \gsim 7.1 \times 10^{-3}$ is ruled out from \cite{Yamazaki:1984vg}.
However, the calculation of the loop-induced $A-s-d$ vertex, relevant for
$K^+ \to \pi^+ + A$, was performed
in \cite{Dolan:2014ska} without a charged Higgs boson in the loops leading
to Ultra-Violet (UV)  divergencies $\sim \ln^2\left(\Lambda/m_{top}\right)$, a factor assumed to
be of ${\cal O}(10)$. As discussed in \cite{Dolan:2014ska}, the divergencies are cancelled
in UV  complete models featuring a light pseudoscalar and in which the
combined contributions to the $A-s-d$ vertex can potentially be much smaller.

An example is provided by the similar process $B \to K +A$ depending on the loop
induced $A-b-s$ vertex, studied in models of the two-Higgs-doublet (+ singlet) type
in \cite{Hall:1981bc,Frere:1981cc,Freytsis:2009ct}. As it can seen in \cite{Freytsis:2009ct}
the partial width can vanish for appropriate choices of parameters (for $M_{H^\pm}\sim
600$~GeV in two-Higgs-doublet models) due to cancellations in the loop functions.
Up to different quark masses, the same loop functions appear in contributions to the
$A-s-d$ vertex. Also within supersymmetric extensions of the SM the
{\it a priori} larger loop contributions to the $A-s-d$ vertex \cite{Hiller:2004ii} can cancel for,
e.g., appropriate values of $A_{top}$ and squark masses within the NMSSM
\cite{Domingo:2016unq}. We estimate that tunings at the 10\% level within
two-Higgs-doublet (+ singlet) models, but at most at the 1\% level within supersymmetric
extensions of the SM would be necessary in order to circumvent
the upper bounds on $\xi_d$ from $K^+ \to \pi^+ + A$. Albeit not elegant, the possibilities
of such cancellations provide a go-theorem allowing for a light pseudoscalar to
circumvent constraints from flavour changing processes in general.

Constraints from searches for $K^+ \to \pi^+ + invisible$ from E787 and BNL-E949
\cite{Adler:2004hp,Artamonov:2009sz} apply only if $A$ decays outside the detectors,
i.e., if $\xi_e$ is small enough. According to \cite{Andreas:2010ms}, identifying
now $C_{Aff}$ in \cite{Andreas:2010ms} with $\xi_e$, this is not the case for $\xi_e \gsim 0.3$.

According to \cite{Dolan:2014ska}, the constraints from $B_s\to \mu^+\mu^-$
(through an off-shell $A$) rule out $g_Y \gsim 0.5$ or $\xi \gsim 0.7$ which is
weaker than the constraint \eqref{boundxid} from $K^+ \to \pi^+ + A$.
Again, the loop contributions to the $A-s-b$ vertex considered in \cite{Dolan:2014ska}
are incomplete within a UV complete extension of the Higgs sector, and
could again be cancelled by additional beyond-the-SM contributions as
in the case of the $A-s-d$ vertex.

The constraints from $B^0\to K^0_S + invisible$ measured by CLEO \cite{Ammar:2001gi}
apply only if the pseudoscalar $A$ produced in $B^0\to K^0_S + A$ decays outside
the detector. Accordingly these constraints depend both on the $Br(B^0\to K^0_S + A)$,
hence on the $A-s-b$ vertex or on $\xi_u,\xi_d$, and on the $A$ decay length which depends on
$\xi_e$. These quantities are identified in \cite{Dolan:2014ska} where a limit
$g_Y \gsim 5$ or $\xi \gsim 3.5$ satisfies the constraints, since then the $A$ decay
length becomes short enough despite the large production rate. Using this constraint
only for $\xi_e$ is conservative, if $\xi_u,\xi_d < \xi_e$ is assumed.

Finally, $\xi_e \gsim 3.5$ satisfies also bounds on $A$ production in radiative $\Upsilon$ decays
$\Upsilon \to \gamma + invisible$ interpreted as $\Upsilon \to \gamma + A$
from CLEO \cite{Balest:1994ch} and BaBar \cite{Aubert:2008as}, which apply only if
$A$ decays outside the detectors. For $M_A\sim 17$~MeV, following \cite{Andreas:2010ms},
this is not the case for $\xi_e \gsim 1.5$.

A second class of constraints on light pseudoscalars originates from beam dump
experiments, which we discuss in turn. First, an electron beam dump on lead experiment
was conducted in Orsay \cite{Davier:1989wz} with the aim to search for light scalar or
pseudoscalar Higgs bosons in the decay into $e^+ e^-$, produced via radiation off electrons.
Correspondingly the resulting constraint applies to $\xi_e$ only.
According to \cite{Davier:1989wz} life times $\tau_A$
in the range $5\cdot 10^{-12}\ \text{s} \lsim \tau_A \lsim 2\cdot 10^{-9}$~s
are ruled out for $M_A\sim 17-18$~MeV. This has already been translated into constraints
on a reduced pseudoscalar-fermion Yukawa coupling $C_{Aff}$ in \cite{Andreas:2010ms},
where $C_{Aff} = \xi_e$ in our notation. Following \cite{Andreas:2010ms},
$0.4 \lsim C_{Aff} \lsim 4$ is ruled out by this constraint.
Since $\xi_e < 0.4$ is incompatible with \eqref{xiegt1}, one is left with
\beq
\label{orsay}
 \xi_e \stackrel{!}{\gsim} 4\; .
\eeq
This constraint leads automatically to the satisfaction of the lower bound
$\xi_e \gsim 3.5$ from $B^0\to K^0_S +\ invisible$, as well as to a short enough
decay length \eqref{xiegt1} for the Atomki pair spectrometer experiment.

Another potentially relevant experiment is the proton beam dump on copper CHARM
experiment \cite{Bergsma:1985qz}. In \cite{Bergsma:1985qz} constraints were
derived assuming that the production cross section and decay length of light
pseudoscalars correspond to the one of axions, which is not the case here.
Relevant is the analysis in \cite{Dolan:2014ska} which uses the production
of light pseudoscalars in $K\to \pi + A$ and $B \to X + A$ decays. For
universally rescaled Yukawa couplings the region $g_Y \gsim 1.5$ or $\xi \gsim 1$
satisfies the constraints, since then the decay length of $A$ is too short to
reach the decay region of the CHARM experiment. This
constraint does not supersede the one in \eqref{orsay}.

The electron beam dump experiment E137 at SLAC \cite{Bjorken:1988as} was
analysed in terms of a decay constant $F$ of leptophilic pseudo-Nambu-Goldstone
bosons in \cite{Essig:2010gu}. From \cite{Essig:2010gu}
one finds that $F \lsim 100$~GeV is allowed which corresponds, with $\frac{1}{F} =
\frac{\xi}{v}$, to $\xi \gsim 2.5$ leading again to a short decay length. Again
this constraint does not supersede the one in \eqref{orsay}.

Constraints from the additional electron beam dump experiments SLAC E141
\cite{Riordan:1987aw} and Fermilab E774 \cite{Bross:1989mp} do not apply for
$M_A \sim 17$~MeV.

Since beam dump experiments are not sensitive to short decay lengths/large
couplings by construction one may ask whether there are any
upper limits on $\xi_e$. Tree level processes mediated by $A$
with Higgs-like Yukawa couplings (even if rescaled by $\xi_e \gsim 4$)
compete with flavour conserving electroweak processes with couplings of
${\cal O}(1)$. Compared to pure electromagnetic processes at eV scales
its contributions are suppressed additionally by (eV/$M_A$)$^4$. Whereas
weak upper limits on $\xi_e$ could certainly be derived from tree level processes,
it is thus not astonishing that presently discussed limits on Yukawa couplings
of $A$ \cite{Dolan:2014ska,Andreas:2010ms,Essig:2010gu} rely on loop-induced
flavour changing processes (and the muon anomalous moment). However, in all
these cases additional BSM particles must contribute in order to
restore electroweak gauge invariance. Since these can cancel the $A$-contribution
for any $\xi_e$ in principle, the upper limit on $\xi_e$ depends on the
amount of finetuning one is willing to tolerate which depends, however, on the
UV-complete model under consideration.

\section{Summary and conclusions}

We studied for which range of Yukawa couplings -- parametrized in terms of
rescaled Yukawa couplings of a SM Higgs boson -- a pseudoscalar with
a mass of $\sim 17$~MeV can explain the anomaly observed in the Atomki pair
spectrometer experiment. The production rate relative to photon emission
in $^8{Be^*}$ decays was estimated in the nuclear shell model (neglecting,
amongst others, isospin--breaking effects) leading to
$\xi_u + \xi_d \approx 0.6$; a larger value is likely if isospin--breaking effects
as discussed in \cite{Feng:2016ysn} are important. A decay length short enough 
for the Atomki pair spectrometer experiment requires $\xi_e \gsim 1$.

Such a light pseudoscalar can generate flavour changing neutral currents which
are constrained notably by $K\to \pi +X$ decays. Here cancellations among
the various (model dependent) loop contributions to the $A-s-d$ vertex,
at least at the 10\% level, must be assumed.
The dominant constraint on $\xi_e$ is $\xi_e \gsim 4$
from the electron beam dump experiment \cite{Davier:1989wz}.

Light pseudoscalars can appear in models with extended Higgs sectors (including
singlets) in which an approximate ungauged global symmetry is spontaneously broken.
Examples are two-Higgs-doublet models of type II with a singlet as the NMSSM
near the Peccei-Quinn or $R$-symmetry limit, in which case one obtains
$\xi_d \sim \xi_e$. On the one hand, given the quite irrevocable constraints on $\xi_e$,
this relation could only be maintained if our result for $\xi_u + \xi_d$
is misleading by an order of magnitude due
to the neglect of isospin breaking, which is not excluded.
On the other hand, larger values for $\xi_d$ would aggravate the required
tuning to suppress $K\to \pi +A$ decays. If these conditions are satisfied,
models for light pseudoscalars from extended Higgs sectors could explain
the anomaly observed in the Atomki pair spectrometer experiment.

\section*{Acknowledgements}

U.~E. and S.~M. acknowledge support from
the grant H2020-MSCA-RISE-2014 No. 645722 (NonMinimalHiggs).
U.~E. acknowledges support from the European Union Initial
Training Networks Higgs\-Tools (PITN-GA-2012-316704), INVISIBLES (PITN-GA-2011-289442),
and the ERC advanced grant Higgs@LHC. He is also grateful to the  University of Southampton for partial financial 
support provided by a  Diamond Jubilee Fellowship.
S.~M. acknowledges partial financial contributions from the NExT Institute and the STFC Consolidated Grant ST/L000296/1.

\end{document}